\documentclass
[english,twocolumn,tightenlines,showpacs,footinbib,groupedaddress,aps,pra]{revtex4}%
\usepackage{amsfonts}
\usepackage{amsmath}
\usepackage{amssymb}
\usepackage{graphicx}
\usepackage{dcolumn}
\usepackage{bm}
\usepackage[usenames]{color}%

\begin{document}
\title{Cavity QED treatment of scattering-induced efficient free-space excitation and
collection in high-Q whispering-gallery microcavities}
\author{Yong-Chun Liu}
\author{Yun-Feng Xiao}
\altaffiliation{To whom correspondence should be addressed. \\
Electronic address:yfxiao@pku.edu.cn.\\
URL: www.phy.pku.edu.cn/$\sim$yfxiao/index.html}

\author{Xue-Feng Jiang}
\author{Bei-Bei Li}
\author{Yan Li}
\author{Qihuang Gong}
\email{qhgong@pku.edu.cn}
\affiliation{State Key Lab for Mesoscopic Physics, School of Physics, Peking University,
Beijing 100871, People's Republic of China}
\date{\today }

\begin{abstract}
Whispering-gallery microcavity laser possesses ultralow threshold, whereas
convenient free-space optical excitation and collection suffer from low
efficiencies due to its rotational symmetry. Here we analytically study a
three-dimensional microsphere coupled to a nano-sized scatterer in the framework of quantum optics. It is found that the scatterer is capable of coupling
light in and out of the whispering-gallery modes (WGMs) without seriously
degrading their high-Q properties, while the microsphere itself plays the role
of a lens to focus the input beam on the scatterer and vice versa. Our
analytical results show that (1) the high-Q WGMs can be excited in free space, and (2) over $50\%$ of the microcavity laser
emission can be collected within less than ${1}^{\circ}$. This coupling system
holds great potential for low threshold microlasers free of external couplers.

\end{abstract}

\maketitle

\section{Introduction}

Whispering-gallery mode (WGM) microcavities represent one of the most
promising candidates for a wide range of fundamental studies and applications,
including cavity quantum electrodynamics, cavity optomechanics, microlasers,
filters and biological sensors (for reviews, see
\cite{Vahala03Rev,WGMRev06,WGMRev10,WGMRev11}). Unfortunately, due to the
rotational symmetry, they suffer from inefficient coupling with the outside
modes, which limits their applications, especially in microlasers
\cite{Laser92APL}. One of the solutions is the tapered fiber coupling method
\cite{taper97OL,taper00PRL,taper03PRL}, which possesses nearly unity
efficiency. Nevertheless, convenient free-space coupling without near-field
couplers is eagerly required because of the experimental limitations
\cite{YSPark06NL,YSPark09NatPhys,pump10PRL}. For example, the external
couplers are not convenient at low temperature chambers; for a
higher-index-material resonator \cite{highn04PRL,highn06PRA}, its coupling
with the tapered fiber is inefficient due to the phase mismatch.
Alternatively, deformed cavities (also named as asymmetric resonant cavity,
ARC) are proposed
\cite{APL93,PRL95,Stone97Nat,ARC98Sci,Lee02PRL,PRL03,PRL03-2,APL03,APL05,PRA05,APL06,NJP06,Xiao09OL,Zou09,PRL10,Shu11PRA,LPR11}%
, because they allow high-efficiency free-space excitation and directional
emission. Latest developments in ARC studies include Limacon-shaped cavity
\cite{limacon08PRL,limacon09APL,limacon09APL2,limacon10PRL} and circular disk
cavity with a linear \cite{ldefect04PRB} or point defect in it
\cite{defect06PRA,defect09PRA}, but the emission divergence angles are still
too broad. Very recently, highly directional outputs are obtained in an
elliptical microdisk with a notch at the boundary \cite{Wang10PNAS}, and by
placing a nanoparticle into the evanescent wave region of microcavities
\cite{Song11OL}. These investigations focused on two-dimensional (2D)
microcavity systems by resorting to numerical simulations.

However, 2D microcavities exhibit relatively low quality ($Q$) factors in
experiment and there are significant energy losses in the perpendicular
dimension. Thus, there is an interest to employ 3D microcavities, whereas it
is difficult to perform numerical simulations. Here we present a cavity
quantum electrodynamics (QED) treatment of 3D microsphere-scatterer coupling
system, and derive analytical expression of the free-space excitation and the
emission directionality. Our results explicitly reveal the underlying physics
and is also suitable for 2D case.

The paper is organized as follows. In Sec II, we briefly describe the
microsphere-scatterer coupling system. In Sec. III and IV, we investigate the
free-space excitation of WGMs and free-space collection of
scattered lasing modes, respectively. Conclusions are presented in Sec. IV.

\section{Scatterer-microsphere system}

\begin{figure}[tb]
\centerline{\includegraphics[width=7.5cm]{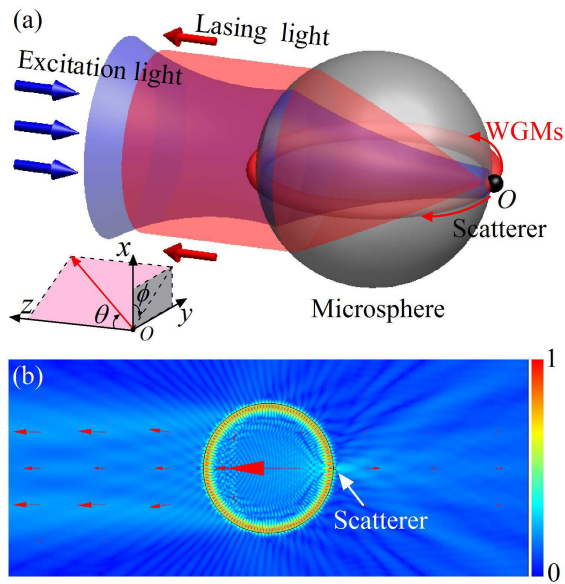}}
\caption{(Color online) (a) Schematic illustration of the
scatterer-microsphere coupling system (not to scale). Bottom left: the corresponding
Cartesian and spherical coordinate systems, with $\theta$ the emitting angle
and $\phi$ the azimuth angle. (b) Finite element simulations (COMSOL Multiphysics) of the
directional emission mode. The color represents the square root of the
electric field's absolute value $\vert E \vert^{1/2}$ (For clarity, we use $\vert E \vert^{1/2}$ in order to increase the contrast). The red arrows indicate the direction of
power flow. In the simulation we use the radii $R=10$ $\mathrm{\mu m}$,
$r_{\text{\textrm{s}}}=200$ \textrm{nm}, microsphere's refractive index
$n=1.7$. }%
\label{fig1}%
\end{figure}

Figure \ref{fig1}(a) illustrates a schematic of the present system. A
spherical subwavelength scatterer (radius $r_{\mathrm{s}}$) locates on the
surface of a microsphere (radius $R$) which is doped with gain medium for the
microlaser applications. A Gaussian pump laser beam (vacuum wavelength
$\lambda_{\mathrm{1}}$, satisfying $r_{\mathrm{s}}\ll\lambda_{\mathrm{1}}\ll
R$) with the polarization in $x$ axis direction propagates along $-z$ axis,
and is incident to the microsphere. Here we have established the Cartesian and
spherical coordinate systems with the scatterer located at the origin $O$, as
sketched in the bottom left of Fig. \ref{fig1}(a). The subwavelength scatterer can be treated as
a dipole \cite{Mazzei07,PRA09}, with the dipole moment induced by the electric
fields of the input Gaussian modes, the excitation and lasing WGMs and the
reservoir modes in the free space. The Rayleigh scattering results in the
interaction among these modes, by which the input photons are scattered into
the excitation WGMs, and the lasing photons in WGMs are scattered into the
reservoir modes. The scattered lasing photons are collimated by the
microsphere, giving rise to directional emission. In Fig. \ref{fig1}(b) the simulation results of wave approach is presented,
demonstrating this directional emission mode.
As a universal paradigmatic
approach, the quantum treatment of Rayleigh scattering is
widely used and well demonstrated
\cite{Mazzei07,Kippenberg09PRL,Zhu10,Yi11PRA,Liu11PRA}. This quantum treatment
is necessary because in such a scatterer-microsphere coupling system there
exists a phenomenon similar with the Purcell effect, since the high state
density of WGMs causes the enhanced scattering between the WGMs and the
free-space reservoir modes \cite{Mazzei07,Kippenberg09PRL}. In the following
we adopt this quantum approach to analyze the free-space excitation of WGMs
and the free-space collection of scattered lasing modes, respectively.

\section{Free-space excitation}

In this section, we develop the theoretical model to describe
the scattering-induced coupling among the surrounding optical modes. By
writing down the total Hamiltonian, we derive the equations of motion (quantum
Langevin equations) for the free-space excitation process. Then we find analytical expressions for the coupling
coefficients and the excitation efficiency. We also address the focusing effect of the microsphere, which
ensures efficient free-space excitation. At last specific examples are presented.

\subsection{General Hamiltonian and Quantum Langevin equations}

Under the rotating wave approximation, the total Hamiltonian\ for the
free-space excitation process can be written as
\begin{align}
H_{\mathrm{1}}^{\text{\textrm{tot}}}  &  =H_{\mathrm{cav}}^{\mathrm{f}%
}+H_{\mathrm{in}}^{\mathrm{f}}+H_{\mathrm{res}}^{\mathrm{f}}\nonumber\\
&  +H_{\mathrm{cav-cav}}^{\mathrm{i}}+H_{\mathrm{cav-in}}^{\mathrm{i}%
}+H_{\mathrm{cav-res}}^{\mathrm{i}}. \label{Htot}%
\end{align}
Here the superscript f (i) labels free (interaction) terms. The first three
terms ($\hslash=1$)
\begin{gather}
H_{\mathrm{cav}}^{\mathrm{f}}=\sum\limits_{m}{\omega_{\mathrm{1,}m}a_{m}%
^{\dag}a_{m},}\label{Hcav}\\
H_{\mathrm{in}}^{\mathrm{f}}=\int\limits_{-\infty}^{+\infty}\omega b^{\dag
}(\omega)b(\omega)d\omega,\\
H_{\mathrm{res}}^{\mathrm{f}}=\sum\limits_{j}{\omega_{j}c_{j}^{\dag}c_{j}},
\end{gather}
describes the free radiation parts, where $a_{m}$, $b(\omega)$ and $c_{j}$
denote annihilation operators of the $m$-th excitation WGM, the input modes
and the $j$-th reservoir mode, with frequencies ${\omega_{\mathrm{1,}m}}$,
$\omega$ and $\omega_{j}$, the commutation relation $[a_{m},{a_{m^{\prime}%
}^{\dag}}]=\delta_{m,m^{\prime}}$, $[b(\omega){,b^{\dag}(\omega^{\prime}%
)]}=\delta(\omega-\omega^{\prime})$, $[c_{j},{c_{j^{\prime}}^{\dag}}%
]=\delta_{j,j^{\prime}}$, respectively. For the WGMs in Eq. (\ref{Hcav}), we
can neglect high order WGMs and focus on the fundamental WGMs, since they
distribute around the equator of the microsphere, possess the minimum mode
volume, the maximum $Q$ factor, and are the typical lasing modes in actual
experiments. Also, In the system described in Fig. \ref{fig1}(a), only
transverse electric (TE) WGMs can be efficiently excited, and transverse
magnetic (TM) WGMs can be safely neglected since the electric field of the TM
WGMs are almost orthogonal to that of the input modes. Therefore, in Eq.
(\ref{Hcav}) the summation index $m$ runs through clockwise (\textrm{CW}) and
counterclockwise (\textrm{CCW}) propagating fundamental WGMs, with a
degenerate frequency denoted as ${\omega_{\mathrm{1,}m}\equiv\omega
_{\mathrm{1}}}$.

The interaction terms of Eq. (\ref{Htot}) are given by
\begin{gather}
H_{\mathrm{cav-cav}}^{\mathrm{i}}=\sum\limits_{m,m^{\prime}}{g_{\mathrm{1}%
,m,m^{\prime}}a_{m}^{\dag}a_{m^{\prime}},}\\
H_{\mathrm{cav-in}}^{\mathrm{i}}={{\sum\limits_{m}\int\limits_{-\infty
}^{+\infty}}}\left[  {{i{{{{g_{\mathrm{in,}m}}}(\omega)b^{\dag}(\omega)a_{m}%
+}}}}\text{ }\mathrm{H.c.}\right]  {d\omega,}\\
H_{\mathrm{cav-res}}^{\mathrm{i}}={\sum\limits_{m,j}({{{g_{\text{\textrm{1}%
},m,j}a_{m}^{\dag}c{_{j}}+\mathrm{H.c.}).}}}}%
\end{gather}
Here $H_{\mathrm{cav-cav}}^{\mathrm{i}}$ describes the interaction between the
cavity modes, which results in the scattering into the same ($m=m^{\prime}$)
or the counterpropagating ($m\neq m^{\prime}$) WGMs with amplitude coupling
strengths ${g_{\mathrm{1},m,m^{\prime}}}$; $H_{\mathrm{cav-in}}^{\mathrm{i}}$
represents the scattering between the input modes and the WGMs with
coefficients ${{{{{{g_{\mathrm{in,}m}}}(\omega)}}}}$; $H_{\mathrm{cav-res}%
}^{\mathrm{i}}$ describes WGMs-reservoir scattering with coefficients
${{{{g_{\text{\textrm{1}},m,j}}}}}$. Without loss of
generality, in the following the coupling coefficients ${g_{\mathrm{1}%
,m,m^{\prime }}}$, ${{{{{{g_{\mathrm{in,}m}}}(\omega }}}})$ and ${g_{\mathrm{1},m,j}}$ are assumed
to be real for notational convenience. The degenerate
\textrm{CW} and \textrm{CCW} WGMs have the same coupling strength in the
scattering process, and thereby we denote ${g_{\mathrm{1},m,m^{\prime}}%
\equiv-g_{\mathrm{1}}}$, ${{{{{{g_{\mathrm{in,}m}}}(\omega}}}})\equiv
{g}{{{{{{_{\mathrm{in}}}}(\omega}}}})$ and ${{{{g_{\text{\textrm{1}},m,j}}}}%
}\equiv{{{{g_{\text{\textrm{1}},j}}}}}$. Note that in Eq. (\ref{Htot}) we have
neglected interacting terms like $H_{\mathrm{in-in}}^{\mathrm{i}}$,
$H_{\mathrm{res-res}}^{\mathrm{i}}$, $H_{\mathrm{in-res}}^{\mathrm{i}}$, since
they do not directly affect the cavity modes, and have minor effect on the
system dynamics.

The eigenmodes of the system are the symmetric and antisymmetric standing
modes, given by $a_{\pm}=(a_{\mathrm{CW}}\pm a_{\mathrm{CCW}})/\sqrt{2}$.
Using the Heisenberg equations and the Markov approximation
\cite{Gardiner,Walls}, we obtain (see Appendix A for details)
\begin{align}
\frac{{da_{+}}}{{dt}}  &  =-i({\omega_{\mathrm{1}}-2g}_{\mathrm{1}}%
{)a_{+}-(\kappa_{\mathrm{in}}+\kappa_{\mathrm{R}})a_{+}}\nonumber\\
&  -\sqrt{2\kappa_{\mathrm{in}}}b_{\mathrm{in}}-\xi, \label{dapdt}%
\end{align}
where
\begin{align}
\kappa_{\mathrm{in}}  &  =2\pi g_{\mathrm{in}}^{2}({\omega_{\mathrm{1}}%
-2g}_{\mathrm{1}}),\\
\kappa_{\mathrm{R}}  &  =2\pi\sum\nolimits_{j}g_{\mathrm{1,}j}^{2}%
\delta(\omega-\omega_{\mathrm{1}}+2g_{\mathrm{1}})
\end{align}
denotes the input-WGMs energy coupling strength and the decay of the WGMs
induced by the Rayleigh scattering to the reservoir. $b_{\mathrm{in}}$ is the
input field and $\xi$ is the noise operator relate to the reservoir.

\subsection{Scattering coefficients and excitation efficiency}

In the full quantum theory, ${{{{{g}_{\mathrm{in}}(\omega}}%
}})$ and ${g_{\mathrm{1}}}$ can be calculated as follows. The quantized
electric field of the excitation WGMs at position $\mathbf{R}$ is given by
$\mathbf{E}_{\mathrm{1}}(\mathbf{R})=[E_{\mathrm{1}}^{(+)}(\mathbf{R}%
)+E_{\mathrm{1}}^{(-)}(\mathbf{R})]\mathbf{\hat{e}}_{x}$. Here $\mathbf{\hat
{e}}_{x}$ is the unit vector along the $x$-axis direction,
\begin{equation}
E_{\mathrm{1}}^{(+)}(\mathbf{R})=\sqrt{\frac{{\hbar\omega}_{\mathrm{1}}%
}{{2\varepsilon_{\mathrm{0}}\varepsilon_{\mathrm{c}}V_{\mathrm{1}}}}%
}f_{\mathrm{1}}(\mathbf{R})(a_{\mathrm{CW}}e^{i\mathbf{k}_{\mathrm{1}%
}\mathbf{\cdot R}}+a_{\mathrm{CCW}}e^{-i\mathbf{k}_{\mathrm{1}}\mathbf{\cdot
R}})
\end{equation}
is the positive frequency component of the field and $E_{\mathrm{1}}%
^{(-)}(\mathbf{R})\ $is its adjoint; $\mathbf{k}_{\mathrm{1}}$ is the wave
vector of the \textrm{CW} mode; ${\varepsilon_{\mathrm{0}}}$ is the dielectric
permittivity of the vacuum and ${\varepsilon_{\mathrm{c}}}$ denotes the
relative permittivity of the microsphere;
\begin{equation}
{V_{\mathrm{1}}=}\frac{\int{\varepsilon}(\mathbf{R})\left\vert E_{\mathrm{1}%
}(\mathbf{R})\right\vert ^{2}d\mathbf{R}^{3}}{\max[{\varepsilon}%
(\mathbf{R})\left\vert E_{\mathrm{1}}(\mathbf{R})\right\vert ^{2}]}%
\end{equation}
is the mode volume of the WGMs, which can be calculated as \cite{WGM89PLA}
\begin{equation}
{V_{\mathrm{1}}=}3.4\pi^{\frac{3}{2}}\left(  \frac{\lambda_{\mathrm{1}}}%
{2\pi{n}}\right)  ^{\frac{7}{6}}R^{\frac{11}{6}}.
\end{equation}
for a microsphere; $f_{\mathrm{1}}(\mathbf{R})=\left\vert E_{\mathrm{1}%
}(\mathbf{R})/E_{\mathrm{1,\max}}\right\vert $ is the normalized field
distribution function of the WGMs. The quantized electric field of the input
beam at position $(\mathbf{r},z)$ is given by $\mathbf{E}_{\mathrm{in}%
}(\mathbf{r},z)=[E_{\mathrm{in}}^{(+)}(\mathbf{r},z)+E_{\mathrm{in}}%
^{(-)}(\mathbf{r},z)]\mathbf{\hat{e}}_{x}$, where $\mathbf{r}=(x\mathbf{\hat
{e}}_{x},y\mathbf{\hat{e}}_{y})$. The positive frequency component reads
\begin{equation}
E_{\mathrm{in}}^{(+)}(\mathbf{r},z)=-i\sum_{k}\sqrt{\frac
{{\hbar\omega}_{k}}{{2\varepsilon_{\mathrm{0}}V}_{k}}}f_{\mathrm{in,}%
k}(\mathbf{r})b_{k}e^{ikz},
\end{equation}
where ${\varepsilon_{\mathrm{b}}}$ denotes the relative permittivity of the
surrounding medium, $b_{k}$ is the annihilation operator of the $k$-th mode,
${V}_{k}$ ($f_{\mathrm{in,}k}(\mathbf{r})$) is the corresponding mode volume
(field distribution function). The expression can be rewritten as an integral
form \cite{Walls}
\begin{equation}
E_{\mathrm{in}}^{(+)}(\mathbf{r},z)=-i\int d\omega\sqrt{\frac{{\hbar\omega}%
}{{4\pi\varepsilon_{\mathrm{0}}cA(z)}}}f_{\mathrm{in}}(\mathbf{r}%
)b(\omega)e^{ikz}.
\end{equation}
Here $c$ is the speed of light in vacuum; ${A(z)}$ is the mode area on $x-y$
plane, given by
\begin{equation}
{A(z)=}\frac{\int\varepsilon(\mathbf{r})\left\vert E_{\mathrm{in}}%
(\mathbf{r})\right\vert ^{2}d\mathbf{r}^{2}}{\max[\varepsilon(\mathbf{r}%
)\left\vert E_{\mathrm{in}}(\mathbf{r})\right\vert ^{2}]}.
\end{equation}
For Gaussian beam ${A(z)}=\iint\nolimits_{-\infty}^{\infty}e^{-2(x^{2}%
+y^{2})/w(z)^{2}}dxdy=\pi w(z)^{2}/2$, where $w(z)$ is the spot radius at $z$.
The quantized electric field of the reservoir is given by $\mathbf{E}%
_{j}(\mathbf{R})=[E_{j}^{(+)}(\mathbf{R})+E_{j}^{(-)}(\mathbf{R}%
)]\mathbf{\hat{e}}_{j}$, where
\begin{equation}
E_{j}^{(+)}(\mathbf{R})=\sqrt{\frac{{\hbar\omega}_{j}}{{2\varepsilon
_{\mathrm{0}}V}_{j}}}a_{j}e^{i\mathbf{k}_{j}\mathbf{\cdot R}},
\end{equation}
with ${V}_{j}$, $\mathbf{k}_{j}$ and $\mathbf{\hat{e}}_{j}$ being the mode
volume, wave vector and the unit vector along polarization direction of the
$j$-th reservoir mode.

The interaction via scattering yields the Hamiltonian
\cite{Mazzei07,Kippenberg09PRL,Zhu10,Liu11PRA}
\begin{equation}
H^{\mathrm{i}}=-\frac{1}{2}\mathbf{p}_{\mathrm{s}}\cdot\mathbf{E}_{\mathrm{s}%
},
\end{equation}
where
\begin{gather}
\mathbf{E}_{\mathrm{s}}=\mathbf{E}_{\mathrm{1}}(\mathbf{0})+\mathbf{E}%
_{\mathrm{in}}(\mathbf{0},0)+\mathbf{E}_{j}(\mathbf{0}),\\
\mathbf{p}_{\mathrm{s}}={\varepsilon}_{\mathrm{0}}\alpha\mathbf{E}%
_{\mathrm{s}}%
\end{gather}
are the total electric field at the position of the scatterer (the origin $O$
of the coordinate system) and the total polarization of the scatterer;
$\alpha=4\pi r_{\mathrm{s}}^{3}(\varepsilon_{\mathrm{s}}-{1})/(\varepsilon
_{\mathrm{s}}+2)$ is the polarizability of the spherical scatterer with
$\varepsilon_{\mathrm{s}}$ being its relative permittivity. Note that for
Gaussian beam the maximum electric field is at the center of the beam,
yielding $f_{\mathrm{in}}(\mathbf{0})=1$. Using the above equations, we
obtain the coupling coefficients as
\begin{gather}
g_{\mathrm{1}}=\frac{\alpha{\omega}_{\mathrm{1}}f_{\mathrm{1}}^{2}%
(\mathbf{0})}{2{\varepsilon_{\mathrm{c}}V_{\mathrm{1}}}},\\
g_{\mathrm{in}}(\omega)=-\frac{1}{2}\alpha f_{\mathrm{1}}(\mathbf{0}%
)\sqrt{\frac{{\omega}_{\mathrm{1}}{\omega}}{{2\pi\varepsilon_{\mathrm{c}%
}cV_{\mathrm{1}}A}_{\mathrm{s}}}},\\
g_{\mathrm{1},j}=-\frac{1}{2}\alpha f_{\mathrm{c}}(\mathbf{0})\sqrt
{\frac{{\omega}_{\mathrm{l}}{\omega}_{j}}{{\varepsilon_{\mathrm{c}%
}V_{\mathrm{l}}V}_{j}}}(\mathbf{\hat{e}}_{x}\cdot\mathbf{\hat{e}}_{j}),
\end{gather}
and thereby the in-coupling strength can be obtained as
\begin{gather}
\kappa_{\mathrm{in}}=\frac{\alpha^{2}f_{\mathrm{1}}^{2}(\mathbf{0}){\omega
}_{\mathrm{1}}({\omega_{\mathrm{1}}-2g}_{\text{\textrm{1}}})}{4{\varepsilon
_{\mathrm{c}}cV_{\mathrm{1}}A}_{\mathrm{s}}},\label{kin}\\
{\kappa_{\mathrm{R}}}=\frac{(n^{5}+1)\alpha^{2}{\omega}_{\mathrm{1}}%
({\omega_{\mathrm{1}}-2g_{\mathrm{1}}})^{3}f_{\mathrm{1}}^{2}(\mathbf{0}%
)}{12\pi c^{3}{\varepsilon_{\mathrm{c}}V_{\mathrm{1}}}}, \label{kR}%
\end{gather}
where ${A}_{\mathrm{s}}=\pi w_{\mathrm{s}}^{2}/2$, with $w_{\mathrm{s}}$ being
the spot radius at $z=0$ plane (where the scatterer is located at).

The excitation efficiency can be defined as $\eta=1-T_{\min}$, where $T_{\min
}\ $is the minimum value of the transmission. From Eq. (\ref{dapdt}), we
obtain (see Appendix B for details)
\begin{equation}
\eta=\frac{4\kappa_{\mathrm{in}}(\kappa_{\mathrm{0}}+\kappa_{\mathrm{R}}%
)}{(\kappa_{\mathrm{0}}+\kappa_{\mathrm{in}}+\kappa_{\mathrm{R}})^{2}}.
\end{equation}

\subsection{Focusing effect of the microsphere and specific examples}

\begin{figure}[tb]
\centerline{\includegraphics[width=6cm]{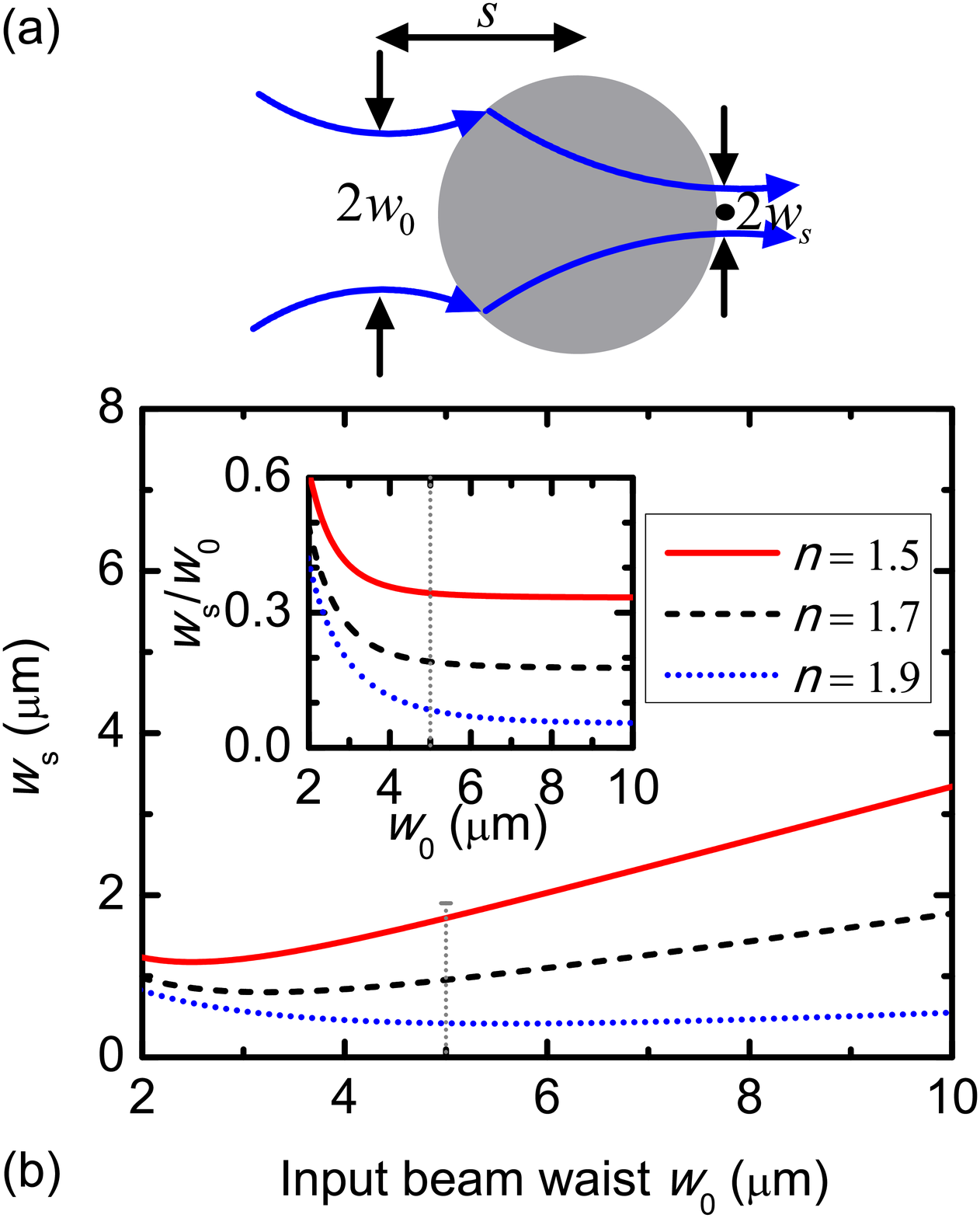}}
\caption{(Color online) (a) Illustration of the microsphere-induced focusing
effect, with $w_{\mathrm{0}}$ being the input beam waist and $w_{\mathrm{s}}$
being the resulting spot radius at $z=0$ plane, after being focused by the microsphere. (b)
$w_{\mathrm{s}}$ vs. $w_{\mathrm{0}}$ for different cavity refractive index
$n$. Inset: spot radius minification $w_{\mathrm{s}}/w_{\mathrm{0}}$ as a
function of $w_{\mathrm{0}}$. The dotted vertical lines indicate
$w_{\mathrm{0}}=5$ $\mathrm{\mu m}$. Here the microsphere's radius $R=10$ $\mathrm{\mu m}$.}%
\label{fig2}%
\end{figure}

To realize efficient coupling, the input beam should be focused into a small
spot on the scatterer, since $\kappa_{\mathrm{in}}$ is in inverse proportion
to the mode area ${A}_{\mathrm{s}}$, as shown in Eq. (\ref{kin}). In fact, the microsphere is a natural
optical lens which possesses ultrashort focal length, capable of focusing the
light spot significantly. Numerical simulation shows that circular dielectric
cylinders illuminated by a plane wave can generate nanojets with waists
smaller than the diffraction limit \cite{jet04OE,jet05OE}. Here we use
Gaussian beam input and analytically treat the problem using Gaussian beam
transform laws. The sphere can be viewed as a thick lens, with the focal
length
\begin{equation}
F=\frac{nR}{2(n-1)},
\end{equation}
where $n$ is the relative refractive index between the sphere and the
environment ($n=\sqrt{{\varepsilon_{\mathrm{c}}}}$). As depicted in Fig.
\ref{fig2}(a), the input light is assumed to be Gaussian beam with a waist
radius $w_{\mathrm{0}}$ ($w_{\mathrm{0}}<R$), and the distance between the
beam waist and the center of the microsphere is $s$. After being focused by
the sphere, the beam waist becomes
\begin{equation}
w_{\mathrm{0}}^{\prime}=\frac{Fw_{\mathrm{0}}}{\sqrt{(s-F)^{2}+z_{\mathrm{R}%
}^{2}}},
\end{equation}
with its distance to the center of the microsphere given by
\begin{equation}
s^{\prime}=\frac{s(s-F)+z_{\mathrm{R}}^{2}}{(s-F)^{2}+z_{\mathrm{R}}^{2}}F,
\end{equation}
where $z_{\mathrm{R}}\equiv\pi w_{\mathrm{0}}^{2}n/\lambda$ is the Rayleigh
range. Then the spot radius\ at $z=0$ plane is
\begin{equation}
w_{\mathrm{s}}=w_{\mathrm{0}}^{\prime}\sqrt{1+\left(\frac{s^{\prime}%
-R}{z_{\mathrm{R}}^{\prime}}\right)^{2}},
\end{equation}
with $z_{\mathrm{R}}^{\prime}\equiv\pi w_{\mathrm{0}}^{\prime2}n/\lambda$.

In Fig. \ref{fig2}(b) we give a specific example, where $R=10$ $\mathrm{\mu
m}$, $s=0$; for dopant Er$^{3+}$, the excitation wavelength $\lambda
_{\mathrm{1}}=977$ \textrm{nm} (lasing wavelength $\lambda_{\mathrm{2}}=1550$
\textrm{nm}). We plot the resulting spot radius $w_{\mathrm{s}}$ as a function
of the input beam waist $w_{\mathrm{0}}$ for different refractive index
$n=1.5$, $1.7$, $1.9$. It is shown that for relatively large input beam waist,
the microsphere is able to focus the beam intensely, especially when $n$
approaches $2$. For $F\ll z_{\mathrm{R}}$, i. e., $w_{\mathrm{0}}^{2}%
\gg\lambda R/[2\pi(n-1)]$, we obtain $w_{\mathrm{s}}\simeq(\left\vert
2-n\right\vert /n)w_{\mathrm{0}}$, which indicates a $\left\vert
2-n\right\vert /n$ times decrease of the sport radius, as further plotted in
the inset of Fig. \ref{fig2}(b). This is consistent with the ray optics
predictions. Therefore, it is of great advantage to make use of the cavity
itself as a micro-sized lens.

\begin{figure}[tb]
\centerline{\includegraphics[width=8.3cm]{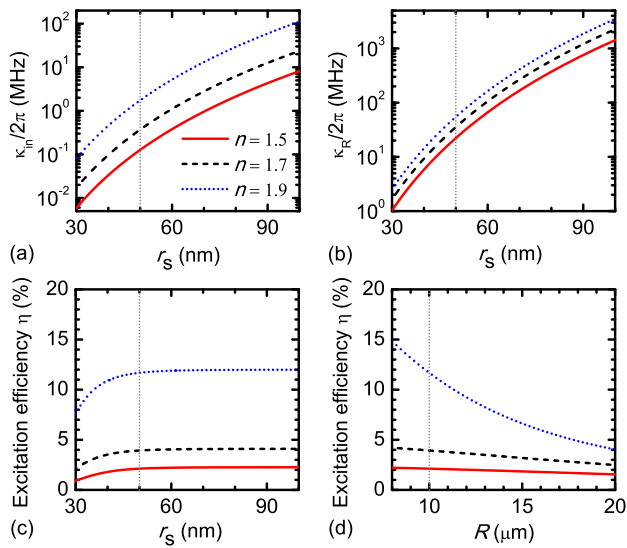}}
\caption{(Color online) (a)-(b) In-coupling strength $\kappa_{\mathrm{in}}$
and Rayleigh scattering induced decay $\kappa_{\mathrm{R}}$ as a function of
the scatterer's radius $r_{\mathrm{s}}$ for different cavity refractive index
$n$. (c)-(d) Excitation efficiency $\eta$ as a function of $r_{\mathrm{s}}$
and $R$. In (a)-(c), we use $R=10$ $\mathrm{\mu m}$, corresponding to the
dotted vertical line in (d); in (d), $r_{\mathrm{s}}=50$ $\mathrm{\mu m}$,
corresponding to the dotted vertical line in (a)-(c).}%
\label{fig3}%
\end{figure}

Now we study the in-coupling strength $\kappa_{\mathrm{in}}$ and Rayleigh
scattering induced decay $\kappa_{\mathrm{R}}$ as a function of the
scatterer's radius $r_{\mathrm{s}}$ as shown in Fig. \ref{fig3}(a)-(b), where
we have set $w_{\mathrm{0}}=5$ $\mathrm{\mu m}$, and other parameters:
$\varepsilon_{\mathrm{s}}=12$ (silicon), $f_{\mathrm{1}}(\mathbf{0})=0.4$ (We will use
there parameters unless specified). The
free-space excitation efficiency for different radii of both the scatterer and
the microsphere are presented in Fig. \ref{fig3}(c)-(d), which shows more than
$10\%$ excitation efficiency can be obtained for suitable parameters. In Fig.
\ref{fig3}(c), for small scatterer, the cavity intrinsic decay $\kappa
_{\mathrm{0}}$ dominates over $\kappa_{\mathrm{in}}$ and $\kappa_{\mathrm{R}}%
$, which results in low excitation efficiency. For large scatterer,
$\kappa_{\mathrm{0}}$ can be neglected compared with $\kappa_{\mathrm{R}}$,
yielding a constant excitation efficiency decided by $\kappa_{\mathrm{in}}$ and $\kappa
_{\mathrm{R}}$. In Fig. \ref{fig3}(d), smaller microspheres possess smaller
mode volumes, resulting in larger $\kappa_{\mathrm{in}}$ and $\kappa
_{\mathrm{R}}$, and thereby larger excitation coefficient. In addition,
smaller microspheres have stronger focusing effects, leading to larger
$\kappa_{\mathrm{in}}$. To obtain a better excitation efficiency, we should
increase $\kappa_{\mathrm{in}}$ and meanwhile decrease $\kappa_{\mathrm{0}}$ and $\kappa_{\mathrm{R}}$.
This can be realized by using smaller input beam waist and by using microcavity with proper
refractive index, which lead to small mode area ${A}_{\mathrm{s}}$, as discussed above.

\section{Free-space collection}

For free-space collection process, the Hamiltonian is similar with that of
free-space excitation process (Eq. (\ref{Htot})) by dropping the terms
containing the input modes. Mention that the energy scattered from the lasing
WGMs into the reservoir modes is just the output of laser emission. Quite
different from the ordinary case in which this kind of scattering leads to
pure damping and is always harmful, here it is a kind of useful resource and
plays a key role in obtaining directional laser emission. {This scattering}
offers an interface between the WGMs inside the microcavity and the optical
modes outside the cavity, where we use the coefficient ${\kappa_{\mathrm{out}%
}}$ to denote the out-coupling strength. Following the calculation of
scattering coefficients in the above section, we obtain
\begin{align}
{\kappa_{\mathrm{out}}}  & =2\pi\sum\nolimits_{j}g_{\text{\textrm{2}%
}\mathrm{,}j}^{2}\delta(\omega-\omega_{\mathrm{2}}+2g_{\mathrm{2}})\\
& =\frac{(n^{5}+1)\alpha^{2}{\omega}_{\mathrm{2}}({\omega_{\mathrm{2}%
}-2g_{\mathrm{2}}})^{3}f_{\mathrm{2}}^{2}(\mathbf{0})}{12\pi c^{3}%
{\varepsilon_{\mathrm{c}}V_{\mathrm{2}}}},
\end{align}
where
\begin{gather}
g_{\mathrm{2}}=\frac{\alpha{\omega}_{\mathrm{2}}f_{\mathrm{2}}^{2}%
(\mathbf{0})}{2{\varepsilon_{\mathrm{c}}V_{\mathrm{2}}}},\\
g_{\mathrm{2},j}=-\frac{1}{2}\alpha f_{\mathrm{2}}(\mathbf{0})\sqrt
{\frac{{\omega}_{\mathrm{2}}{\omega}_{j}}{{\varepsilon_{\mathrm{c}%
}V_{\mathrm{2}}V}_{j}}}(\mathbf{\hat{e}}_{x}\cdot\mathbf{\hat{e}}%
_{j}),\label{g2j}%
\end{gather}
and ${V_{\mathrm{2}}}$ ($f_{\mathrm{2}}(\mathbf{0})$) is the mode volume
(field distribution function) of the lasing WGMs.

\begin{figure}[tb]
\centerline{\includegraphics[width=6cm]{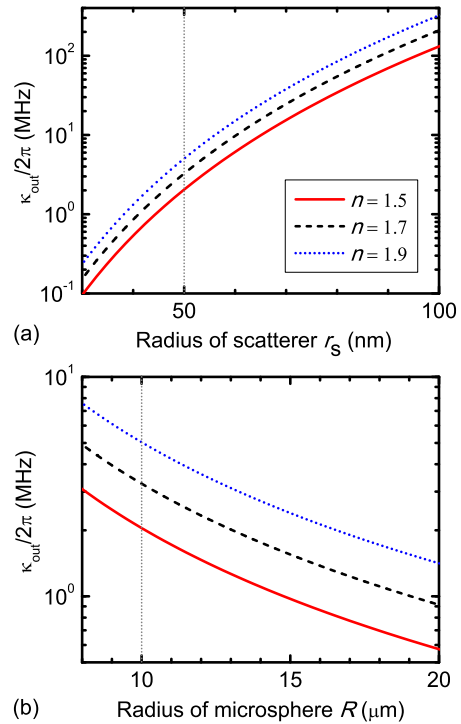}}
\caption{(Color online) Out-coupling strength $\kappa_{\mathrm{out}}$ as a
functions of the scatterer's radius $r_{\mathrm{s}}$ (a) and microsphere's
radius $R$ (b) for different cavity refractive index $n$.}%
\label{fig4}%
\end{figure}

Figure \ref{fig4} plots ${\kappa_{\mathrm{out}}}$ as a function of the radii
of the scatterer and the microsphere. For $r_{\mathrm{s}}=50$ \textrm{nm} and
$R=10$ $\mathrm{\mu m}$, ${\kappa_{\mathrm{out}}}$ is several mega Hertz,
which corresponds to $Q_{\mathrm{out}}={\omega}_{\mathrm{2}}/{\kappa
_{\mathrm{out}} \sim10}^{8}$. Therefore, the high-$Q$ properties of the lasing
modes can be maintained.

In the following we analyze the emission directionality originating from
Rayleigh scattering and collimating effect of the microsphere. Finally the
emission directionality and energy collection ratio for various parameters are present.

\subsection{Scattering directionality}

The factor $\mathbf{\hat{e}}_{x}\cdot\mathbf{\hat{e}}_{j}$ in Eq. (\ref{g2j}) indicates
that $g_{\mathrm{2},j}$ depends on the direction, resulting the
direction-dependent out-coupling coefficient ${\kappa(\theta,\phi)}$, which
satisfies ${\kappa_{\mathrm{out}}=}\iint{\kappa(\theta,\phi)}d\Omega$, with
$\Omega$ being the solid angle and $d\Omega=\sin{\theta d\theta d\phi}$. After
normalization, we can define ${u(\theta,\phi)=\kappa(\theta,\phi
)/\kappa_{\mathrm{out}}}$, which can be calculated as
\begin{equation}
u(\theta,\phi)=%
\begin{cases}
\frac{3n^{5}(1-\sin^{2}{\theta}\cos^{2}{\phi})}{{4\pi(n^{5}+1)}}%
,\ \ {0}^{\circ}\leq\theta<{180}^{\circ},\\
\frac{3(1-\sin^{2}{\theta}\cos^{2}{\phi})}{{4\pi(n^{5}+1)}},\ \ {180}^{\circ
}\leq\theta<{360}^{\circ}.
\end{cases}
\end{equation}

\begin{figure}[tb]
\centerline{\includegraphics[width=6cm]{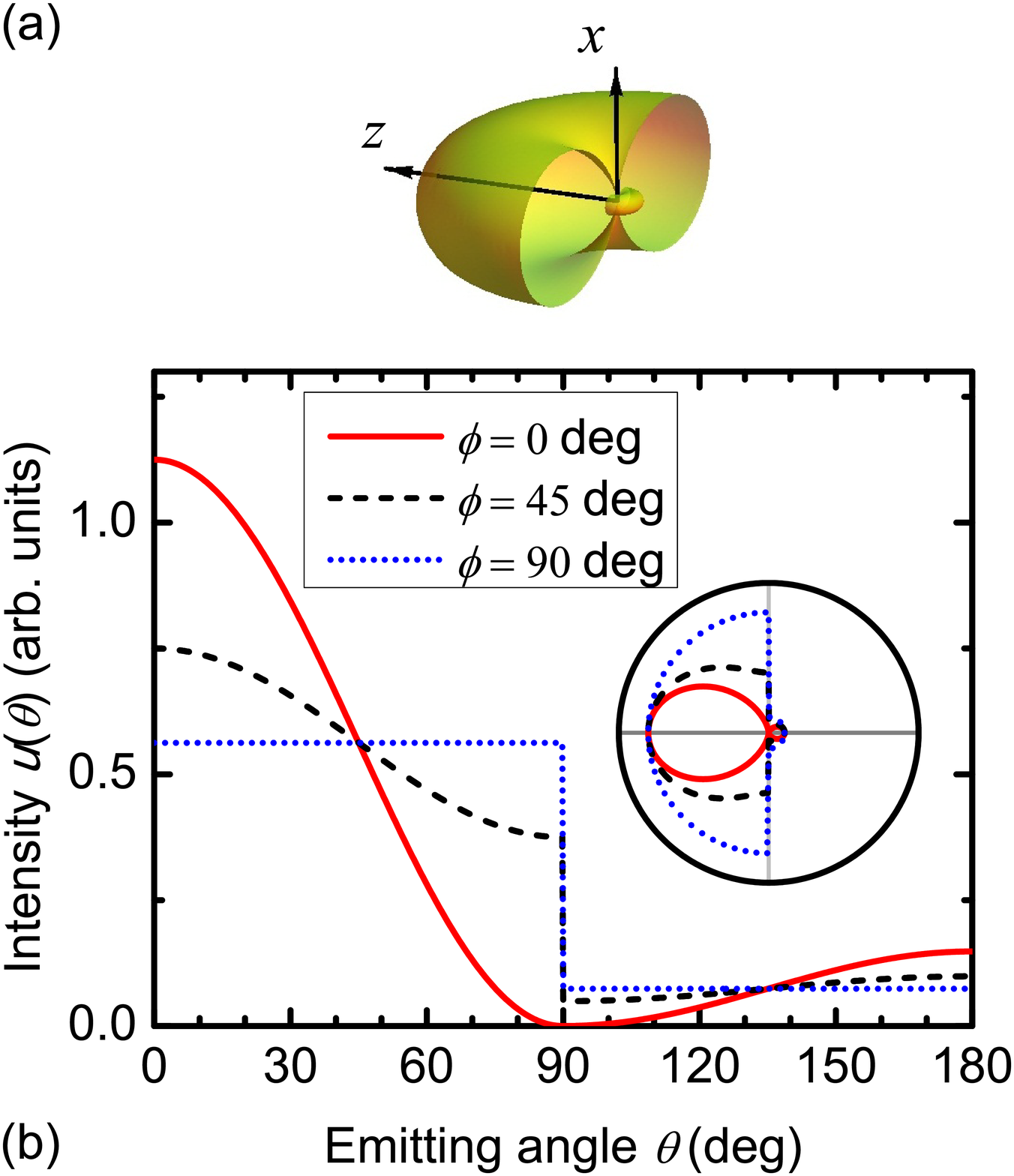}}
\caption{(Color online) (a) Spherical plot of $u(\theta,\phi)$. (b)
$u(\theta,\phi)$ vs. $\theta$ for various $\phi$. Inset: The corresponding
polar plot of $u(\theta,\phi)$. Here we use $n=1.5$.}%
\label{fig5}%
\end{figure}

The normalized out-coupling coefficient ${u(\theta,\phi)}$ represents the
angular distribution of output energy. In Fig. \ref{fig5}(a) we present
${u(\theta,\phi)}$ for given azimuth angle ${\phi=0}^{\circ}$, ${45}^{\circ}$,
${90}^{\circ}$, respectively. Note that for ${0}^{\circ}\leq\theta
<{180}^{\circ}$ the environment is dielectric cavity with permittivity of
$\varepsilon_{\mathrm{c}}$, while for ${180}^{\circ}\leq\theta<{360}^{\circ}$
that is vacuum with permittivity of $1$. As shown in Fig. \ref{fig5}, the
light trends to be scattered to $z$ axis (${\theta=0}^{\circ},18{0}^{\circ})$,
but the scattering along $+z$ axis is much stronger due to the asymmetry
environment. Note that for ${\phi=90}^{\circ}$, the scattering is uniform in
the same environment, since in this case the scattered light has the same
polarization with the WGMs.

\subsection{Collimating effect of the microsphere}

\begin{figure}[tb]
\centerline{\includegraphics[width=6cm]{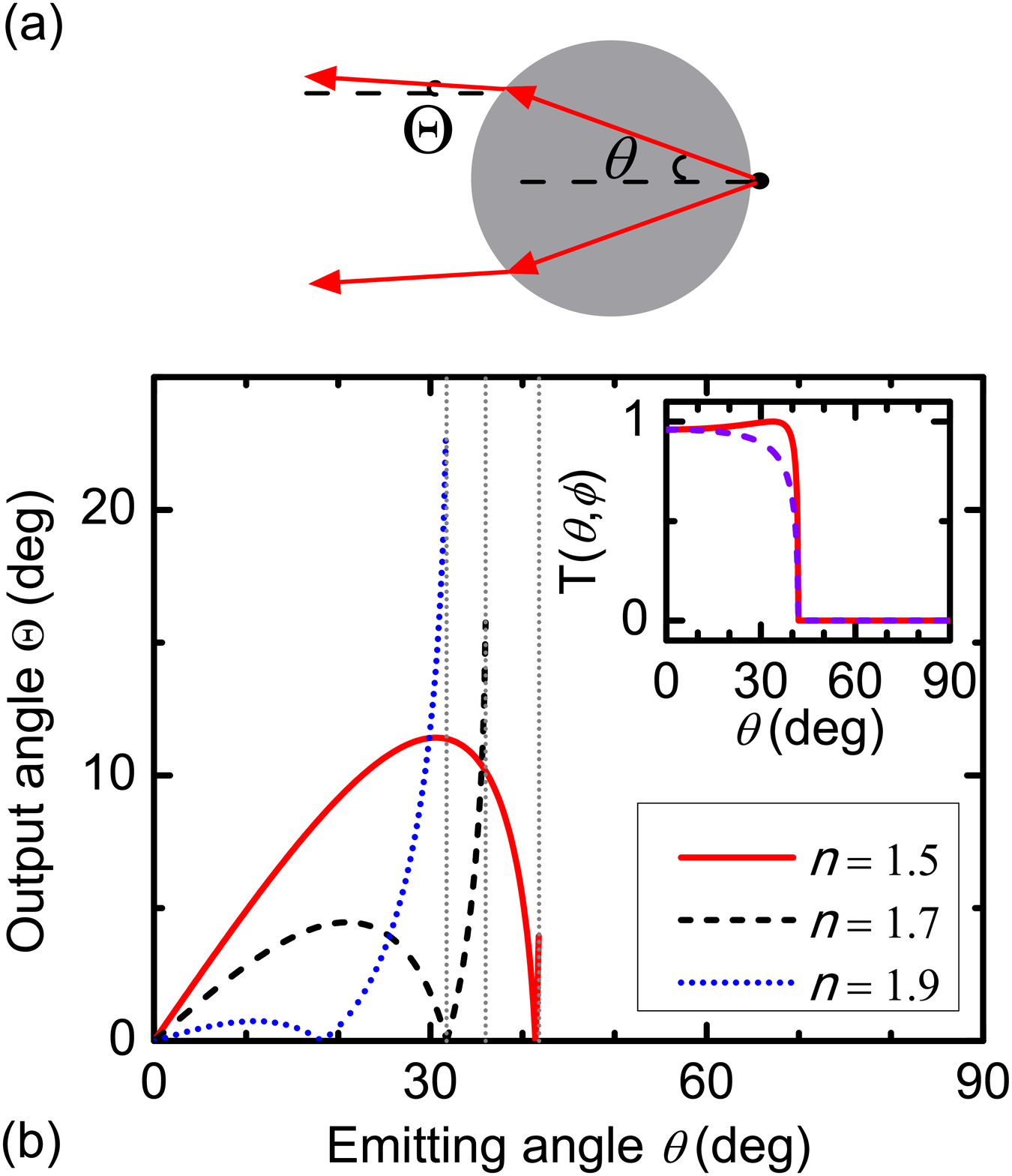}}
\caption{(Color online) (a) Illustration of the microsphere-induced
collimating effect. (b) $\Theta$ vs. $\theta$ for $n=1.5$, $1.7$, $1.9$. The
dotted vertical lines correspond to the critical angles of total reflection
for $n=1.5$, $1.7$, $1.9$ (from right to left). Inset: Transmission as a function of $\theta$ for
$\phi={0}^{\circ}$ (solid), $90^{\circ}$ (dashed), here $n=1.5$.}%
\label{fig6}%
\end{figure}

Although the scattered light trends to propagate along $z$ axis, the
divergency angle is too wide. To obtain better directionality, once again we
can make use of the microsphere itself, which behaves as a thick lens. As
depicted in the inset of Fig. \ref{fig6}(b), the emitted light for
${0}^{\circ}{\leq\theta<{(}18{0}^{\circ}/\pi)\arcsin(1/n)}$ (the critical angles of total reflection) passes trough the
microsphere, and finally yields the output angle $\Theta$, given by
\begin{equation}
\Theta=f({\theta})=\left\vert {2\theta}-\arcsin\left(n\sin{\theta}\right)\right\vert.
\end{equation}
Note that for ${{(}18{0}^{\circ}/\pi)\arcsin(1/n)\leq\theta<90}^{\circ}$, the
light is totally reflected on the microsphere surface. From Fig. \ref{fig6}(b)
we can see that the output angle $\Theta$ has a much small divergence than the
emitting angle ${\theta}$, which stems from the collimating effect of the microsphere.

The output energy density function{\small \ }can be obtained as
\begin{equation}
p(\Theta,\phi)=\sum T(f^{-1}(\Theta),\phi)u(f^{-1}(\Theta),\phi)\frac
{df^{-1}(\Theta)}{d\Theta},
\end{equation}
where{\small \ }%
\begin{equation}
T(\theta,\phi)=T_{p}(\theta,\phi)\cos^{2}\phi+T_{s}(\theta,\phi)\sin^{2}\phi;
\end{equation}
$T_{p}$ ($T_{s}$) is the transmission for p-polarization (s-polarization)
component, calculated from Fresnel formula; ${\theta=f^{-1}(\Theta)}$ is the
inverse function of $f({\theta})$, which is a multiple valued function, and
$\sum$ means the summation over each section of the multiple valued function.
Note that $p(\Theta,{\phi})$ has singularities, thus the
full-width-of-half-maximum definition of the divergence angle fails. To
quantify the emission directionality, we define a half-energy angle
$\Theta_{1/2}$, given by $P(\Theta_{1/2},\phi)=1/2$, where
\begin{equation}
P(\Theta,\phi)=\int_{0}^{\Theta}p(\Theta^{\prime},{\phi})d\Theta^{\prime}%
\end{equation}
is the energy ratio (energy distribution function), representing how much
energy distributes in the interval $[0,\Theta]$. This half-energy angle
represents that the output angle of half light is smaller than $\Theta_{1/2}$.

\begin{figure}[tb]
\centerline{\includegraphics[width=7.5cm]{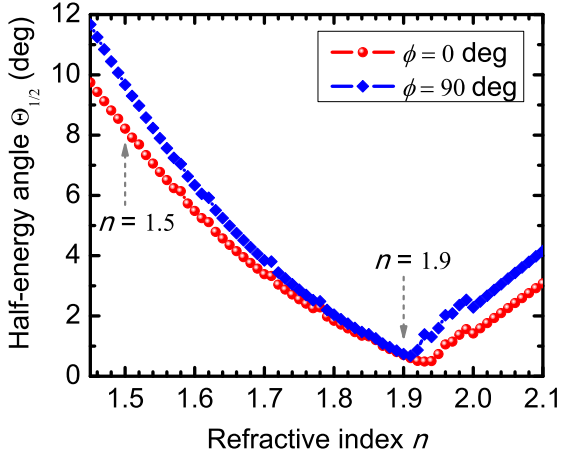}}
\caption{(Color online) (a) Half energy angle $\Theta_{1/2}$ vs. $n$ for
$\phi={0}^{\circ}$, $90^{\circ}$.}%
\label{fig7}%
\end{figure}

In Fig. \ref{fig7} we plot the half-energy angle as a function of the
refractive index for $\phi={0}^{\circ}$ and ${90}^{\circ}$. Note that these
two cases set the lower and upper bounds, as inferred from the emission
pattern in Fig. \ref{fig5}(a). Remarkably, $\Theta_{1/2}$ can be less than
${1}^{\circ}$ (for refractive index around $1.9$), which indicates much better directionality than previous
predictions based on other mechanisms, to the best of our knowledge.

\begin{figure}[tb]
\centerline{\includegraphics[width=8.3cm]{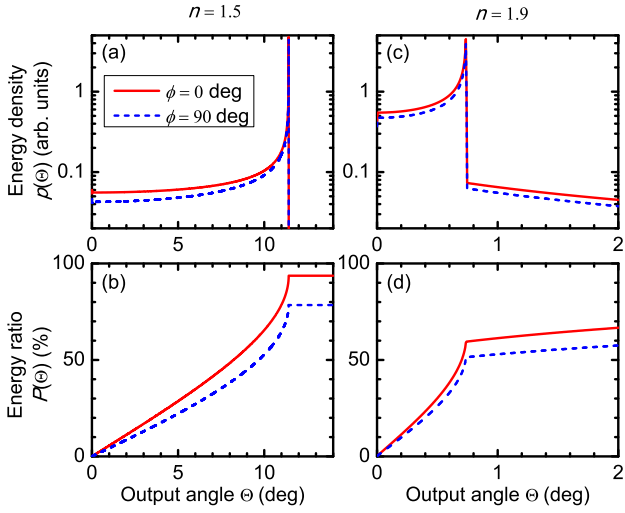}}
\caption{(Color online) Energy density $p(\Theta)$ and energy ratio
$P(\Theta)$ as functions of $\Theta$ different $n$ and $\phi$. (a)-(b):
$n=1.5$; (d)-(e): $n=1.9$. }%
\label{fig8}%
\end{figure}

As two specific cases, the energy density function $p(\Theta,{\phi})$ and the energy ratio $P(\Theta
,\phi)$ for $n=1.5$ and $1.9$ are plotted in Fig. \ref{fig8}(b)-(e). For $n=1.5$, more than
$80\%$ emission energy can be collected within $\Theta\simeq{11}^{\circ}$. For $n=1.9$, very good directionality can be obtained, with
over $50\%$ collection efficiency for $\Theta\leq{0.7}^{\circ}$. 

For widely used microcavities with the refractive index not equal to 1.9 (e.g., Silica, Calcium Fluoride, Lithium Niobate), the optimal emission directionality can be obtained by optimizing the shape of the microcavity (e.g., using deformed cavities) and the position of the scatterer (e.g. embedding the scatterer inside the microcavity). Note that this optimal refractive index is irrelevant with the size, shape and material of the scatterer.

\section{Conclusion}

In summary, based on cavity QED approach, we analytically investigate the
microsphere-scatterer coupling system in which the high-Q WGMs can be
efficiently excited through free space and the resulting laser is capable of
emitting with high directionality. In this system, a subwavelength scatterer
placed in the vicinity of the microsphere serves as an interface between the
input light, the WGMs and the output light. We take advantage of the
microsphere itself to focus the input beam with small spot area, and collimate
the output beam with ultra-small divergence angle. Our results show that
the high-Q WGMs can be excited with efficiency larger than $10\%$. More
importantly, the half-energy angle of the output light can be as narrow as
${0.7}^{\circ}$, which is a great improvement over the 2D microcavity lasers
\cite{Wang10PNAS,Song11OL}. This holds great potential for novel micro-sized
laser sources and has broad applications in Micro/Nano photonics.

\begin{acknowledgments}
This work was supported by the NSFC (Grants No. 10821062, No. 11004003, and
No. 11023003) and the 973 program (Grant No. 2007CB307001).
\end{acknowledgments}

\appendix

\section{Derivation of Quantum Langevin equations}

Starting from the total Hamiltonian $H_{\mathrm{1}}^{\text{\textrm{tot}}}$
(Eq. (\ref{Htot})), the Heisenberg equations of motion can be derived as
\begin{widetext}%
\begin{gather}
\frac{{da_{m}}}{{dt}}=-i{\omega _{\mathrm{1}}a_{m}}-\int\limits_{-\infty
}^{+\infty }{g}{{{{{{_{\mathrm{in}}}}(\omega }}}})b(\omega ){d\omega +ig_{%
\mathrm{1}}\sum\limits_{m^{\prime }}a_{m^{\prime }}-i\sum\limits_{j}{g}{{{{{{%
_{\mathrm{1},j}}}}}c_{j}}}}, \\
\frac{{db(\omega )}}{{dt}}=-i\omega b(\omega )+{g}{{{{{{_{\mathrm{in}}}}%
(\omega }}}}){\sum\limits_{m}a_{m}}, \\
\frac{{dc}_{j}}{{dt}}=-i\omega _{j}{c}_{j}-{i{g}{{{{{{_{\mathrm{1},j}}}}}}}%
\sum\limits_{m}{{a_{m}}}}.
\end{gather}
\end{widetext}

The eigenmodes of the system are the symmetric and antisymmetric standing
modes, given by $a_{\pm}=(a_{\mathrm{CW}}\pm a_{\mathrm{CCW}})/\sqrt{2}$.
Then the above equations can be rewritten as
\begin{widetext}%
\begin{gather}
\frac{{da_{+}}}{{dt}}=-i{\omega _{\mathrm{1}}a_{+}}-\sqrt{2}%
\int\limits_{-\infty }^{+\infty }{g}{{{{{{_{\mathrm{in}}}}(\omega }}}}%
)b(\omega ){d\omega +i2g_{\mathrm{1}}a_{+}-i\sqrt{2}\sum\limits_{j}{g}{{{{{{%
_{\mathrm{1},j}}}}}c_{j}}}},  \label{ap} \\
\frac{{da_{-}}}{{dt}}=-i{\omega _{\mathrm{c}}a_{-}}, \\
\frac{{db(\omega )}}{{dt}}=-i\omega b(\omega )+\sqrt{2}{g}{{{{{{_{\mathrm{%
in}}}}(\omega }}}}){a_{+}},  \label{dbdt} \\
\frac{{dc}_{j}}{{dt}}=-i\omega _{j}{c}_{j}-i{\sqrt{2}{g}{{{{{{_{\mathrm{1}%
,j}}}}}a_{+}}}}.  \label{dcdt}
\end{gather}
\end{widetext}
Formal integrations of Eq. (\ref{dbdt}) and (\ref{dcdt}) yield
\begin{widetext}%
\begin{gather}
b(\omega)=e^{-i\omega (t-t_{0})}b_{\mathrm{0}}(\omega)+\sqrt{2}g_\mathrm{in}(\omega)\int\limits_{t_{0}}^{t}{e^{-i\omega
(t-t^{\prime })}a_{+}(t^{\prime })dt^{\prime },} \\
{c}_{j}=e^{-i\omega _{j}(t-t_{0})}c_{j,\mathrm{0}}-i\sqrt{2}g_{\mathrm{1},j}\int\limits_{t_{0}}^{t}{e^{-i\omega _{j}(t-t^{\prime
})}a_{+}(t^{\prime })dt^{\prime },}
\end{gather}
\end{widetext}
where $b_{\mathrm{0}}(\omega)$, $c_{j,\mathrm{0}}$ denotes the value of
$b(\omega)$, $c_{j}$ at $t=t_{0}$, respectively. In both equations the first
terms represent the free evolution of the modes while the second terms arise
from the interaction with the WGMs.

Substituting the solutions into Eq. (\ref{ap}), we finally obtain
\begin{widetext}%
\begin{gather}
\frac{{da_{+}}}{{dt}}=-i({\omega _{\mathrm{1}}-2g}_{\mathrm{1}}{%
)a_{+}-(\kappa _{\mathrm{in}}+\kappa _{\mathrm{R}})a_{+}}-\sqrt{2\kappa _{%
\mathrm{in}}}b_{\mathrm{in}}-\xi ,  \label{dap}
\end{gather}
\end{widetext}
where
\begin{equation}
\kappa_{\mathrm{in}}=2\pi g_{\mathrm{in}}^{2}({\omega_{\mathrm{1}}%
-2g}_{\mathrm{1}})
\end{equation}
represents the input-WGMs energy coupling strength,
\begin{equation}
b_{\mathrm{in}}=\frac{1}{\sqrt{2\pi}}\int_{-\infty}^{+\infty}b_{\mathrm{0}%
}(\omega)e^{-i\omega(t-t_{0})}d\omega
\end{equation}
describes the input field,
\begin{equation}
{\kappa_{\mathrm{R}}}={2\pi\sum\nolimits_{j}}g_{\mathrm{1,}j}^{2}\delta
(\omega-{\omega_{\mathrm{1}}+2g}_{\mathrm{1}})
\end{equation}
denotes the damping of the WGMs induced by the scattering to the reservoir,
\begin{equation}
\xi={i\sqrt{2}\sum\nolimits_{j}{g}{{{{{{_{\mathrm{1},j}}}}}}}}e^{-i\omega
_{j}(t-t_{0})}c_{j,\mathrm{0}}%
\end{equation}
is the noise operator relate to the reservoir. In deriving Eq. (\ref{dap}), we
have used the Markov approximation \cite{Gardiner}.

\section{Derivation of excitation efficiency}

Taking the intrinsic decay rate $\kappa_{\mathrm{0}}$\ of the WGMs into
account, and using the input-output relation \cite{Gardiner,Walls}
\begin{equation}
b_{\mathrm{out}}=b_{\mathrm{in}}+\sqrt{2\kappa_{\mathrm{in}}}a_{+},
\end{equation}
we obtain
\begin{equation}
b_{\mathrm{out}}=b_{\mathrm{in}}+\frac{2\kappa_{\mathrm{in}}b_{\mathrm{in}%
}+\sqrt{2\kappa_{\mathrm{in}}}\xi}{i(\omega-{\omega_{\mathrm{1}}%
+2g}_{\mathrm{1}}{)-(\frac{\kappa_{\mathrm{0}}}{2}+\kappa_{\mathrm{in}}%
+\kappa_{\mathrm{R}})}}.
\end{equation}
Then the transmission can be obtained as $T=\langle b_{\mathrm{out}%
}^{\dag}b_{\mathrm{out}}\rangle /\langle b_{\mathrm{in}}^{\dag
}b_{\mathrm{in}}\rangle $. For optical frequency, room temperature, the
initial states of the reservoir modes are almost all vacuum states. Thus the
expectation values of the noise operators can be neglected. Therefore, we
obtain the minimum value of the transmission
\begin{equation}
T_{\min}=\left(  \frac{2\kappa_{\mathrm{in}}-\kappa_{\mathrm{0}}%
-2\kappa_{\mathrm{R}}}{2\kappa_{\mathrm{in}}+\kappa_{\mathrm{0}}%
+2\kappa_{\mathrm{R}}}\right)  ^{2}.
\end{equation}
Thus the excitation efficiency $\eta=1-T_{\min}$ can be obtained as
\begin{equation}
\eta=\frac{4\kappa_{\mathrm{in}}(\kappa_{\mathrm{0}}+2\kappa_{\mathrm{R}}%
)}{(\kappa_{\mathrm{0}}+2\kappa_{\mathrm{in}}+2\kappa_{\mathrm{R}})^{2}}.
\end{equation}

\end{document}